\documentclass[a4paper,11pt]{article}
\pdfoutput=1
\usepackage{jheppub}

\usepackage{amsmath, amssymb, mathrsfs}

\usepackage{graphicx}
\usepackage{dcolumn}
\usepackage{bm}
\usepackage{color}
\usepackage{cleveref}

\newcommand{\phid}{\dot{\phi}}

\newcommand{\Hphi}{H_{,\phi}}

\renewcommand{\d}[1]{\mathrm{d}#1}

\renewcommand{\L}{\Lambda}

\title{Hamilton-Jacobi formalism for Generalized Chaplygin Gas models}

\date{\today}

\author[a]{Yordan Ignatov\,,}
\author[a]{Mauro Pieroni\,,}

\affiliation[a]{Theoretical Physics, Blackett Laboratory, Imperial College, London, SW7 2AZ, United Kingdom}

\emailAdd{yi17@imperial.ac.uk}
\emailAdd{m.pieroni@imperial.ac.uk}

\abstract{ In this work we discuss the application of the Hamilton-Jacobi formalism on the scalar field implementation of Generalized Chaplygin Gas models. This corresponds to a Generalised Born-Infeld action for the scalar field, which in an initial fast-rolling phase mimics a matter-like behavior and in the final slow-rolling phase mimics a cosmological constant. In order to enrich the phenomenology of the model, we add an extra functional freedom, specified through a scalar potential for the field. Interestingly, we find that, due to the lifting induced by the non-standard kinetic term, an asymptotic de Sitter-like configuration can be obtained even for negative potentials. We show that at the background level, this model can easily mimic the $\Lambda$CDM model both with and without independent baryonic and radiation components.   }

\begin{document}
	
\maketitle
\flushbottom

\section{Introduction}
\label{sec:intro}

The standard cosmological model known as $\Lambda$CDM, where $\Lambda$ stands for the Cosmological Constant (CC) modeling Dark Energy (DE) and CDM is Cold Dark Matter (CDM), completed by cosmic inflation is widely accepted as the most successful model to describe the evolution of our Universe on large scales. In particular, the remarkable agreement between this model and the most recent Cosmic Microwave Background (CMB) data~\cite{Planck:2018vyg} provides a strong evidence in its favor. However, despite its successes, the $\Lambda$CDM is commonly regarded as not completely satisfactory from the theoretical point of view. For example, the fine tuning problem for the CC originally formulated by Weinberg~\cite{Weinberg:1988cp} poses a question on the foundedness of a CC to model DE. For this reason, over the years several alternative models for DE have been introduced. Among the most famous possibilities it is worth mentioning quintessence~\cite{Peebles:1987ek, Ratra:1987rm, Caldwell:1997ii, Zlatev:1998tr}, K-essence~\cite{Chiba:1999ka, Armendariz-Picon:2000nqq, Armendariz-Picon:2000ulo}, scalar-tensor theories~\cite{Uzan:1999ch, Chiba:1999wt, Amendola:1999qq, Perrotta:1999am, Bertolami:1999dp, Boisseau:2000pr, Esposito-Farese:2000pbo, Deffayet:2011gz, Kobayashi:2011nu}, $f(R)$ theories~\cite{Capozziello:2002rd, Capozziello:2003tk, Carroll:2003wy} and other modifications of gravity~\cite{Nojiri:2006ri}. For reviews of the different possibilites see for example~\cite{Copeland:2006wr, EuclidTheoryWorkingGroup:2012gxx}.\\

An intriguing possibility to simultaneously model DM and DE is to introduce an exotic fluid, which in the original proposal of~\cite{Kamenshchik:2001cp} was dubbed the Chaplygin gas, with a non-standard equation of state: 
\begin{equation}
    \label{eq:GCG_EOS}
    p = -\frac{A}{\rho^\alpha} \; , 
\end{equation}
where $\alpha =1$ and $A$ is a positive parameter. In the generalization proposed in~\cite{Bento:2002ps}, which is customarily dubbed the Generalized Chaplygin Gas (GCG), the parameter space for this model was extended to $ 0 < \alpha \leq 1$. By directly integrating~\cref{eq:GCG_EOS} it is easy to show that the energy density for this fluid reads:
\begin{equation}
    \label{eq:GCG_rho}
    \rho = \left( \frac{B}{a(t)^{3(1+\alpha)}} + A \right)^\frac{1}{1+\alpha} \;,
\end{equation}
where $B > 0$ is an integration constant and $a(t)$ is the scale factor of the Friedmann-Lemaitre-Robertson-Walker (FLRW) metric\footnote{As customary, we have set $a(t_0) = 1$ where $t_0$ is the value of $t$ today.}. From this expression it is manifest that, because of the cosmological growth of the scale factor, the GCG features different behaviours: at early times (implying $a \ll 1$), the first term in~\cref{eq:GCG_rho} dominates and the GCG behaves like CDM. On the other hand, as the Universe (and consistently $a$) grows, the second term dominates and $\rho$ becomes constant, mimicking DE. \\

After the original proposal of~\cite{Kamenshchik:2001cp} (see also~\cite{Bilic:2001cg} which also included a first discussion of cosmological perturbations) and the generalization of~\cite{Bento:2002ps}, several works tried to assess the viability of GCG models~\cite{Gorini:2002kf, Avelino:2002fj, Makler:2002jv, Bento:2002yx, Sandvik:2002jz, Bean:2003ae}. As pointed out in~\cite{Sandvik:2002jz}, strong evidence against this class of models is the presence of oscillatory behaviours in the evolution of the density perturbations. However, as already stated in the conclusions of~\cite{Sandvik:2002jz}, this feature can be avoided if the effective speed of sound differs from the adiabatic sound speed (see for example~\cite{Hu:1998kj}). A first study of the impact of entropy perturbations in a Universe filled with baryons and a GCG was proposed in~\cite{Reis:2003mw} which showed that instabilities and oscillations can disappear in the presence of entropy perturbations. It is also worth mentioning~\cite{Beca:2003an} which has pointed out the importance of including an independent baryon fluid in the analyses involving a GCG. After these first works, many other authors have discussed the viability of GCG models in a fully consistent cosmological setup~\cite{Amendola:2003bz, Zhu:2004aq, Bento:2005un, Wu:2007bv, Gorini:2007ta, Xu:2010zzb, Freitas:2010mw, Lu:2010zzj, Campos:2012ez, Carneiro:2014jza, El-Zant:2015dya, Xu:2016grp, vomMarttens:2017cuz, Aurich:2017lck, Wen:2017aaa}. While a GCG cosmology is slightly disfavoured with respect to standard $\Lambda$CDM, models featuring small values of $\alpha$ (\emph{i.e.} $\alpha \lesssim 0.1$ or smaller which practically mimic a cosmological constant at late time) can still still be considered viable~\footnote{It is worth mentioning that it has recently been claimed~\cite{Yang:2019nhz, DiValentino:2021izs} that these models could ease the tension between the Planck collaboration's~\cite{Planck:2018vyg} and the SH0ES collaboration's~\cite{Riess:2020fzl} measurement of the Hubble constant $H_0$.}. Since an analysis of perturbations lies beyond the scopes of this paper, we will limit our analysis to values of $\alpha$ which are compatible with the constraints given by the most recent works on topic~\cite{Wen:2017aaa}.\\

In this work we discuss the GCG model in terms of the Hamilton-Jacobi (HJ) formalism for cosmology developed by Salopek and Bond~\cite{Salopek:1990jq}. In particular, we consider a slight generalization (due to the introduction of a scalar potential) of the scalar field implementation of a GCG~\cite{Bento:2002ps}. While in the literature these techniques have extensively been applied to both models of inflation and quintessence~\cite{Lidsey:1995np, Hoffman:2000ue, Kinney:2002qn, Liddle:2003py, Binetruy:2014zya, Cicciarella:2016dnv, Cicciarella:2017nls, Thompson:2018ifr}, to the best of our knownledge it is the first time that this approach is used to study GCG cosmologies. As we show in the following, within this framework it is reasonably easy to get an analytical understanding of the evolution of the system. In turn, this insight on the dynamics may provide useful guidelines for model building. \\ 

The paper is structured as follows: in~\cref{sec:model} we discuss the scalar field implementation of GCG and its generalization due to the introduction of a scalar potential. In~\cref{sec:HJ} we express the models in terms of the HJ formalism and we build a suitable parameterization for the Hubble parameter. In~\cref{sec:cosmological} we frame the discussion of the previous sections in a more complete cosmological environment. In~\cref{sec:conclusions} we further discuss our results and we draw our conclusions. In~\cref{sec:appendix_A} we present a scan over the parameter space of our model, showing the role played by each variable.

\section{The model}
\label{sec:model}
A scalar field implementation of the GCG introduced in~\cref{eq:GCG_EOS} and~\cref{eq:GCG_rho} of~\cref{sec:intro} can be obtained from a Generalised Born-Infeld (GBI) theory~\cite{Bento:2002ps} described by the action\footnote{As customary we work in a FLRW Universe, choosing a mostly positive metric signature $(-+++)$. We use Planck units, setting $c=8\pi G_N = 1$.}
\begin{equation}
\label{eq:GCG_action}
    S = \int \d^4 x \sqrt{-g} \left( \frac{R}{2} - A^{\frac{1}{1+\alpha}} \left[1 - (-2 X)^{\frac{1 + \alpha}{2 \alpha }}\right]^{\frac{\alpha }{1+\alpha}} \right) \;,
\end{equation}
where $\phi(t)$ is a homogeneous scalar field and $X \equiv g^{\mu\nu} \partial_\mu \phi \, \partial_\nu \phi/2$. Notice that for $\alpha=1$, and up to a missing $\mp$ factor for $D3/\overline{D3}$-branes respectively, we recover the usual Dirac-Born-Infeld (DBI) action which describes the motion of $D3/\overline{D3}$-branes, so that $A$ can be interpreted as the (squared) warp factor. In the limit of small $X$ (and up to a field redefinition to absorb $A$) this reduces to the usual expression $p(\phi, X) = -X -V(\phi)$ for standard canonically normalized scalars (and with a constant potential expressed in terms of $A$). It is easy to show that Einstein equations for this theory read:
\begin{align}
    3H^2 & =  \rho = A^{\frac{1}{1+\alpha}} \left[ 1-(-2X)^{\frac{1+\alpha}{2 \alpha }}\right]^{-\frac{1}{1+\alpha}}  \;, \label{eq:fried1}  \\
    -2\dot{H} & =  p + \rho = A^{\frac{1}{1+\alpha}} \left[  1 - (-2X)^{\frac{1+\alpha}{ 2 \alpha}} \right]^{-\frac{1}{1+\alpha}} (-2X)^\frac{1+\alpha}{2 \alpha} \label{eq:fried2} \;.
\end{align}
By comparing~\cref{eq:fried1} with \cref{eq:fried2} it is manifest that this theory satisfies the GCG equation of state in~\cref{eq:GCG_EOS}. For completeness we report the expression for the \emph{adiabatic} speed of sound
\begin{equation}
c_s^2 \equiv \left. \frac{\delta p}{\delta \rho}\right|_{\delta \phi = 0} =  \alpha   \left[1 - (-2 X)^{\frac{1 + \alpha}{2 \alpha }}\right]  =  \frac{ \alpha A} { \rho ^{1 +\alpha} } \; , \label{eq:sound_speed}
\end{equation}
so that it is clear that the choice of restricting the parameter space to $0 < \alpha \leq 1 $ is required in order to ensure $0 < c_s^2 \leq 1$.\\

As discussed in the literature~\cite{Amendola:2003bz, Aurich:2017lck, Wen:2017aaa}, a GCG with the addition of a baryonic component can produce a reasonable fit of the cosmological evolution of the Universe. The GBI implementation of the GCG outlined in this section is fully specified by the choice of its kinetic term which, in practice, is achieved by fixing the two parameters $\alpha$ and $A$. Namely, this model has parametric but not functional freedom. In order to further explore the capability of this theory to describe the evolution of a unified dark fluid, we propose a slight generalization of this model which introduces and additional functional freedom. This is can be easily achieved by adding a homogeneous scalar potential $V(\phi)$ for the field $\phi$. With this modification the pressure and energy densities read:
\begin{align}
\label{eq:p_rho_phi}
    p(X, \phi) 
    &= -A^{\frac{1}{1+\alpha}} \left( 1 - (-2X)^{\frac{1+\alpha}{2\alpha}} \right)^{\frac{\alpha}{1+\alpha}} - V(\phi) \;, \\
\label{eq:rho_phi}
    \rho(X, \phi) 
    &= +A^{\frac{1}{1+\alpha}} \left(1-(-2X)^{\frac{1+\alpha}{2 \alpha }}\right)^{-\frac{1}{1+\alpha}} + V(\phi) \;.
\end{align}
which can directly be used to show that the Equation of State (EOS) parameter for this modified implementation of the GCG reads
\begin{equation}
\label{eq:EOS}
    1+w = \frac{p + \rho}{\rho} = \frac{ \left(1-(-2X)^{\frac{1+\alpha}{2 \alpha }}\right)^{- \frac{1}{1+\alpha}} (-2X)^\frac{1+\alpha}{2\alpha} }{  \left(1-(-2X)^{\frac{1+\alpha}{2 \alpha }}\right)^{-\frac{1}{1+\alpha}} + V(\phi) \, A^{ - \frac{1}{1+\alpha}}  } \;.
\end{equation}
In the slow-roll limit of $(-2X) \rightarrow 0 $, and assuming $V(\phi)$ does not go to zero faster than $(-2X)^\frac{1+\alpha}{2\alpha}$,~\cref{eq:EOS} vanishes, implying DE-like behaviour. In the opposite limit of $(-2X) \rightarrow 1$, the kinetic term dominates over the potential in the denominator, resulting in DM-like behaviour. Notice also that by definition, the expression of the adiabatic speed of sound $c_s$ given in~\cref{eq:sound_speed} is not affected by the introduction of the scalar potential. In the following we will refer to this model as the GBI+V model.\\

Before concluding this section, it is worth stressing that for $(-2X) \rightarrow 0$ the non-standard kinetic term of the GBI (and thus of the GBI+V) does not vanish. As a consequence, in the limit $(-2X) \rightarrow 0$ the energy density of the GBI model does not vanish but rather approaches $A^{\frac{1}{1+\alpha}}$. In order to have a better understanding of the dynamics of the GBI+V models it is thus useful to introduce an effective kinetic term $K_{eff}(X)$ and an effective potential $V_{eff}(\phi)$ as:
\begin{equation}
\begin{aligned}
    \label{eq:effective_pot_def}
    \rho(X, \phi) &= A^{\frac{1}{1+\alpha}} \left\{ \left(1-(-2X)^{\frac{1+\alpha}{2 \alpha }}\right)^{-\frac{1}{1+\alpha}} -1 \right\}  + V (\phi) + A^{\frac{1}{1+\alpha}} \;, \\[5pt]
    &\equiv  K_{eff}(X) + V_{eff}(\phi) \; .
\end{aligned}
\end{equation}
Notice that with these definitions $K_{eff}(X) \rightarrow 0 $ as $X \rightarrow 0$ and, after reabsorbing the $A^{\frac{1}{1+\alpha}}$ term, the potential $V_{eff}(\phi)$ is effectively lifted with respect to $V(\phi)$. Interestingly, this allows for the construction of models with negative $V(\phi)$ which can still realize a stable de Sitter (dS) configuration. In fact, for an appropriate choice of the parameters of the model, the lifing induced by the non-standard kinetic term keeps the $V_{eff}(\phi)$ positive, which allows to avoid the big crunch that normally occurrs in presence of negative potentials~\cite{Linde:2001ae, Felder:2002jk, Heard:2002dr}.

\section{Hamilton-Jacobi formalism for unified dark matter/energy}
\label{sec:HJ}
The HJ formalism is based on the assumption that the time evolution of the scalar field $\phi(t)$ is piecewise monotonic, so that it is possible  to get $t(\phi)$ \emph{i.e.} to use the field itself as a clock to describe the evolution of the system. Within this framework, the Hubble parameter can be expressed as a function of $\phi$ as:
\begin{equation}
\label{eq:Hubble}
    H(\phi) \equiv \frac{\d \ln a}{\d t} = \frac{\d \ln a}{\d \phi} \, \phid \;,
\end{equation}
where both $a$ and $\phid$ are also functions of $\phi$ only. In order to fully specify the evolution of the system, we also need Einstein equations~\footnote{From now on, without loss of generality, we work under the assumption that $\phi > 0$ and $\phid > 0$.}:
\begin{align}
    3H^2 &= A^{\frac{1}{1+\alpha}} \left[ 1-\phid^{\frac{1+\alpha}{\alpha }}\right]^{-\frac{1}{1+\alpha}} +V(\phi) \;, \label{eq:HJ_fried1} \\[5pt]
    -2\phid \Hphi &= A^{\frac{1}{1+\alpha}} \left[  1 - \phid^{\frac{1+\alpha}{\alpha}} \right]^{-\frac{1}{1+\alpha}} \phid^\frac{1+\alpha}{\alpha} \;, \label{eq:HJ_fried2}
\end{align}
where we have used the chain rule to express $\dot{H}=\phid H_{\phi}$. As a first step to study the evolution of the system, we can solve~\cref{eq:HJ_fried2} for $\phid$ to get:
\begin{equation}
\label{eq:phid(phi)}
    \phid = \left( 1 + A(-2\Hphi)^{-(1+\alpha)} \right)^{-\frac{\alpha}{1+\alpha}} \;.
\end{equation}
This can then be substituted into~\cref{eq:HJ_fried1} to get a first order differential equation which, for a given choice of $V(\phi)$, can be solved to get the evolution of the Hubble parameter as a function of $\phi$. Notice that in the $V(\phi) = 0$ case, this reduces to
\begin{equation}
   ( 3H^2 )^{1 + \alpha} =  A + (-2\Hphi)^{1 + \alpha}  \; ,
\end{equation}
meaning that the evolution of $H(\phi)$ is uniquely determined by the choice of $\alpha$, $A$ and $\phi_0$ corresponding to the value of $\phi$ today. On the other hand, the addition of a scalar potential provides the system with an extra functional freedom which can be chosen to modify the evolution. In this work we proceed with a slightly different (and more bottom-up) approach. Rather than specifying $V$ and solving a differential equation for $H$, we choose an ansatz for $H$ which algebraically fixes $V$ through~\cref{eq:HJ_fried1}. This not only strongly simplifies the solution of the system but, as we will show in the following, also gives a quite direct analytical understanding of the different limiting behaviours corresponding to DM and DE respectively. \\

\begin{figure}
    \centering
    \includegraphics[width=0.9\textwidth]{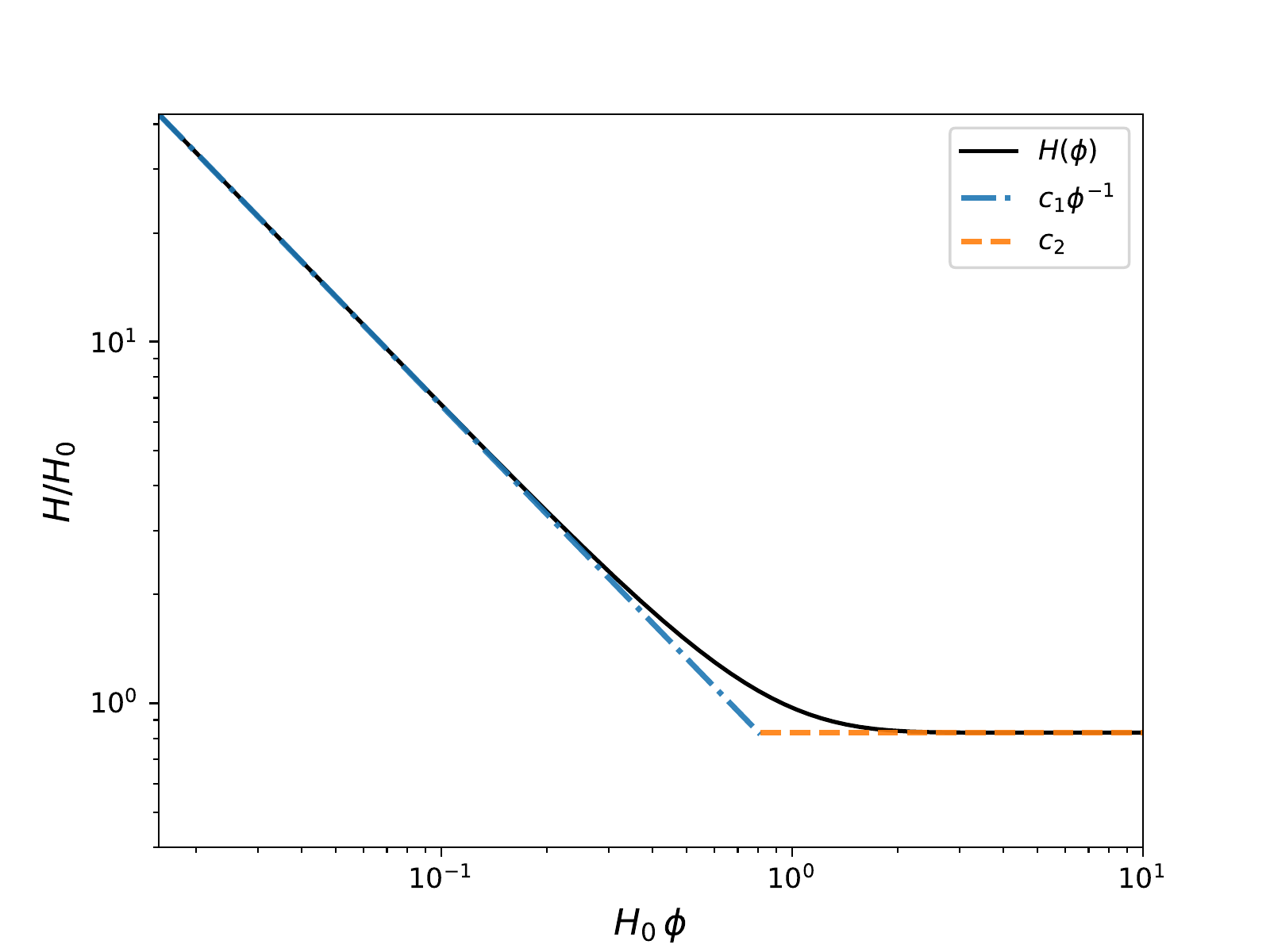}
    \caption{Comparision between the complete numerical evolution of $H(\phi)$ in a GCG model with $\alpha = 0.02$ and $A=(3\Omega_\L H_0^2)^{1+\alpha}$ and the two asymptotic behaviours given in~\cref{eq:Hubble_req}. For this particular model (and in these units) the two prefactors $c_1$ and $c_2$ are adimensional and respectively equal to $2/3$ and $0.83$.}
    \label{fig:H_behaviour}
\end{figure}

Before discussing in detail the explicit form of the parameterization for $H(\phi)$ let us first focus on the $V = 0 $ case. As we show in the following, under this assumption it will be possible to derive simple analytical approximations of the relevant quantities. These approximations will then be used as building blocks to define a suitable expression for $H(\phi)$. Let us start by considering~\cref{eq:EOS} which, for $V = 0$, reduces to:
\begin{equation}
    \label{eq:eos_GCG}
    1+w  = \phid^\frac{1+\alpha}{\alpha} \;.
\end{equation}
This equation (or alternatively~\cref{eq:phid(phi)}) can be directly integrated to get $t(\phi)$ and, using $H \equiv \dot{a}/a$, also $a(\phi)$. We can then consider the early time limit, where we want the GCG to reproduce a matter-like fluid. This corresponds to $w\simeq0$, which implies $\phid \simeq 1$ and thus $t \simeq \phi$. Since for DM domination we have $H \simeq 2/(3t) \propto a^{-3/2}$ we can immediately conclude that, in order for the GCG to reproduce DM at early times, we need both $H \propto \phi^{-1}$ and $a \propto \phi^{2/3}$. On the other hand, DE domination would correspond to a nearly constant Hubble parameter implying that a suitable parameterization for $H$ should smoothly interpolate between the two regimes~\footnote{The $2/3$ prefactor in the DM phases is kept to directly recover the correct limiting behaviour for $a(t)$.}:
\begin{equation}
\label{eq:Hubble_req}
    H(\phi) \propto \left\{ \begin{array}{cc}
        \frac{2}{3}\phi^{-1} \,, &\;\; \mathrm{DM} \; , \\
        const\,, &\;\; \mathrm{DE} \; .
    \end{array}
    \right. \;
\end{equation}
For our purposes, we may ignore the contribution of $\phid$ to the Hubble parameter, since in the dark matter phase the field is quickly rolling, with $\phid$ very close to unity. The two regimes of~\cref{eq:Hubble_req} are shown in~\cref{fig:H_behaviour} in comparison with the evolution of the Hubble parameter for a GCG model which, as expected, smoothly interpolates between them. Here, we have chosen a value of $\alpha=0.02$ as it lies within the accepted range \cite{Wen:2017aaa}, while $A$ is chosen such that the GCG model emulates the behaviour of the $\L$CDM model in the distant future. This is done by equating~\cref{eq:GCG_rho} to the joint energy density of DM and DE in the $\L$CDM model, resulting in $A=(3\Omega_\L H_0^2)^{1+\alpha}$. In practice we use the $H_0=67.37\mathrm{km}\,\mathrm{s}^{-1}\mathrm{Mpc}^{-1}, \Omega_\L=0.6889$ from Planck 2018~\cite{Planck:2018vyg}. Notice that for this choice the cosmological evolution of the GCG closely resembles $\L$CDM.\\

\begin{figure}
\makebox[\textwidth][c]{
    \includegraphics[width=\textwidth]{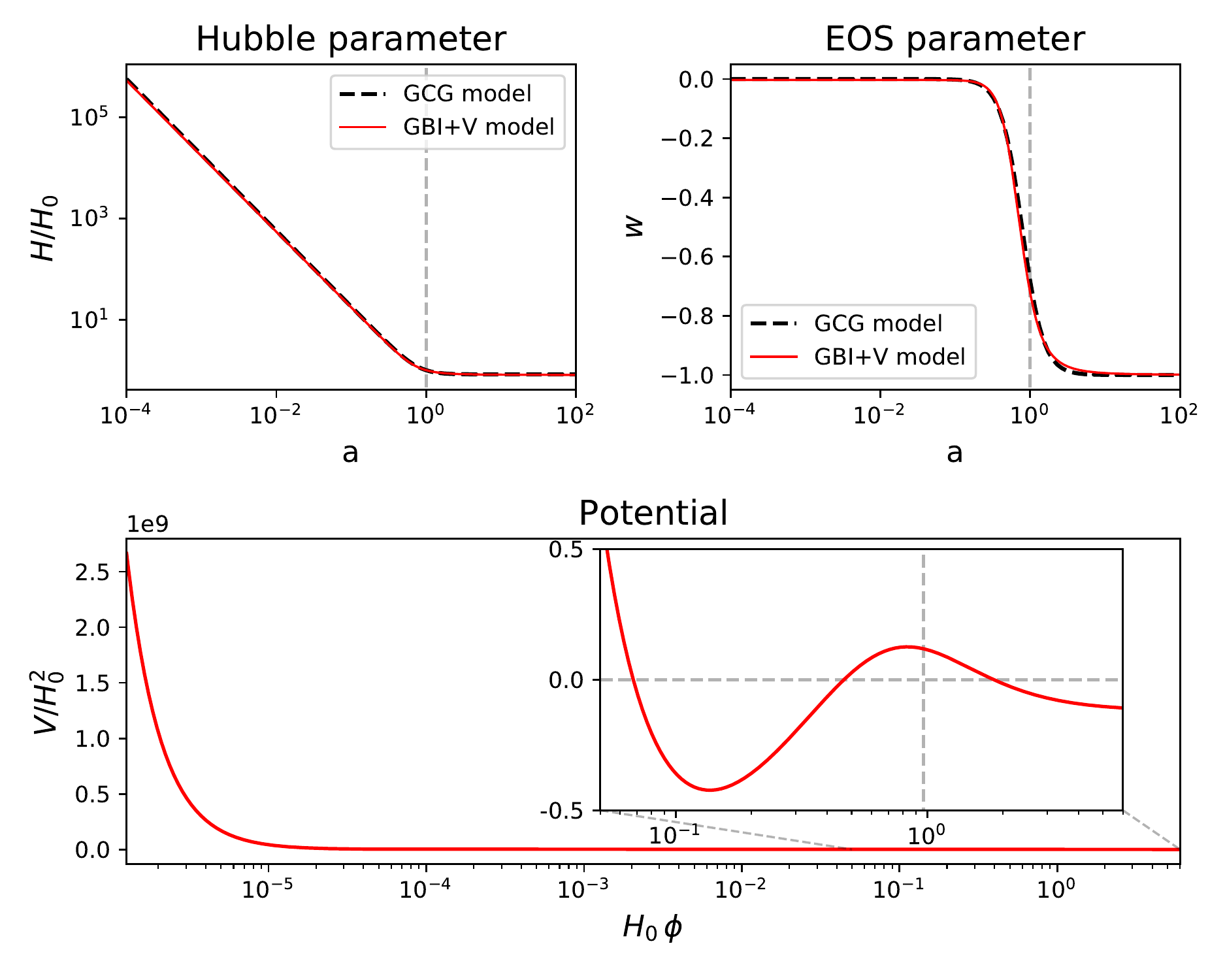}}
    \caption{Top plots: comparison between the GCG model (black dashed line), corresponding to a pure GBI model, and a model respecting~\cref{eq:param} (\emph{i.e.} a GBI+V model) with $\{p_i\} = \{0.669, 0.830, 2.438\}$ (red line). The vertical dashed lines correspond to $a=1$. The top left and right plots respectively show the evolution of $H$ and of the equation of state parameter as a function of $a$. The bottom plot shows the shape of the scalar potential as a function of $\phi$. The field range used in the bottom plot corresponds to the range used for $a$ in the two top plots.}
    \label{fig:scalarGCGpot}
\end{figure}

Guided by the discussion of the previous paragraph, we can then specify a parameterization for $H(\phi)$ which, in the $V \neq 0 $ case, will be suitable to model the transition from DM to DE. In particular we choose:
\begin{equation}
\label{eq:param}
    H(\phi) = \frac{p_1}{\phi} \left(1 + \left(\frac{H_0\,\phi}{p_2}\right)^{p_3}\right)^{\frac{1}{p_3}} \equiv \frac{p_1}{\phi} G(\phi) \;.
\end{equation}
where $H_0$ is the value of the Hubble parameter today, the $\{p_i\},\,i\in \{1, 2, 3\}$ are three positive (and order one) constants and, for later convenience, we introduced the function $G$. This ansatz for $H$ clearly features the two behaviours presented in~\cref{eq:Hubble_req}. For $ \phi H_0 \ll p_2$ (which will correspond to early times) we have $G\rightarrow 1$ and thus $H \simeq p_1/\phi$. On the other hand in the opposite limit (which will correspond to present time and future) we have $ \phi H_0 \gg p_2$ and thus $G\rightarrow \phi H_0/p_2$ implying $H \simeq H_0 p_1 / p_2$ corresponding to an asymptotic dS-like phase. Let us comment on the impact of the three parameters $p_i$: $p_1$ (and in particular it's ratio with $p_2$) directly controls the late time value of $H$ and value of the EOS parameter in the DM-like phase. The second parameter $p_2$ sets the value of $\phi$ corresponding to the DM-DE transition. Finally, $p_3$ controls the sharpness of the transition from DM to DE. In order to get a model which closely resembles the cosmological evolution given by the standard GCG for an appropriate choice of $\alpha$ and $A$, we fit the three $p_i$ parameters to match the evolution of $H(\phi)$ of the GCG model. This procedure gives $\{p_i\}\simeq\{0.669, 0.830, 2.438\}$, which are the parameters we will use for specifying the GBI+V model in the HJ approach. The cosmological evolution of $H$ and of the equation of state parameter (as functions of $a$) for this choice of parameters, are shown in the two top plots of~\cref{fig:scalarGCGpot} in comparison with the GCG model with $\alpha = 0.02$ and $A = (3\Omega_\L H_0^2)^{1+\alpha}$. A more detailed analysis of the parameter space for the parameterization in~\cref{eq:param} and the plots showing the impact of the three parameters on the evolution of the system are shown in~\cref{sec:appendix_A}.\\

To conclude this section, we discuss the explicit form of $V$ given by~\cref{eq:HJ_fried1} once the parameterization of $H$ given by~\cref{eq:param} is set. By substituting~\cref{eq:phid(phi)} into~\cref{eq:HJ_fried1} we can immediately get:
\begin{equation}
    \label{eq:potential}
    V(\phi) =  \frac{1}{\phi^2} \left\{   3p_1^2\,G(\phi)^{ 2}  -  2 p_1  G(\phi)^{1 - p_3}  \left[1 + A\left(\frac{2p_1}{\phi^2}\,G(\phi)^{1 - p_3}\right)^{-(1+\alpha)}\right]^{\frac{1}{1+\alpha}} \right\} \;.
\end{equation}
The shape of the potential for a model with $\{p_i\} = \{0.669, 0.830, 2.438\}$, is shown in the bottom plot of~\cref{fig:scalarGCGpot}. As we show in the following, the left part (\emph{i.e.} before the dip) of the potential is a power law in $\phi$, and it is sufficiently steep to sustain a fast-roll of the scalar field. On the other hand, the right part (\emph{i.e.} after the dip) flattens and asymptotically approaches a constant. The dip appears in correspondence to the transition between this two regimes and induces a quick decrease in the field velocity ensuring a quick transition between the two regimes. As discussed in~\cref{sec:appendix_A}, the height of this dip can be tuned by varing $p_3$. To get a better understanding of the shape of the potentail we proceed by deriving the asymptotic expressions for $V$ in the two limiting behaviours discussed in the previous paragraph. At early times (\emph{i.e.} in DM domination) we have $ \phi H_0  \ll p_2 $ and $G \rightarrow 1 $ so that~\eqref{eq:potential} reduces to:
\begin{equation}
    \label{eq:V_small_phi}
   V(\phi) \simeq \frac{p_1}{\phi^2}  \left( 3p_1\,  - 2 \left[1 + A\left(\frac{2p_1}{\phi^2}\, \right)^{-(1+\alpha)} \right] ^{\frac{1}{1+\alpha}}\right) \propto \frac{p_1}{\phi^2} (3p_1-2)  \;,
\end{equation}
which corresponds to a steep potential to support the initial fast rolling phase giving DM domination. On the other hand in the opposite limit we have: 
\begin{equation}
    \label{eq:future_potential}
    V(\phi)  \simeq   3 \,\left( \frac{p_1 H_0}{p_2}\right)^{ 2} \left[ 1 + \frac{2}{p_3} \left( \frac{H_0 \phi}{p_2} \right)^{-p_3}  \right]- A^\frac{1}{1+\alpha}  \; , 
\end{equation}
which corresponds to an asymptotically flat and lower bounded potential. Notice that in the infinite future (where the $\phi$ dependent terms can be neglected) the potential may be negative if:
\begin{equation}
    3 \left( \frac{p_1 H_0}{p_2} \right)^2  < A^{\frac{1}{1+\alpha}} \; , 
\end{equation}
but, by construction, the Hubble parameter is always positive. As we have already discussed at the end of~\cref{sec:model}, this is due to the kinetically induced lifting of the effective potential introduced in~\cref{eq:effective_pot_def}. In particular for $ \phi H_0  \gg p_2 $ the effective potential reads:
\begin{equation}
    V_{eff}(\phi)  \simeq   3 \,\left( \frac{p_1 H_0}{p_2}\right)^{ 2} \left[ 1 + \frac{2}{p_3} \left( \frac{H_0 \phi}{p_2} \right)^{-p_3}  \right] + A^\frac{1}{1+\alpha}     \left( \frac{H_0 \phi}{p_2} \right)^{-p_3}  \; , 
\end{equation}
which is clearly always positive. This implies that the lifting induced by the non-standard kinetic term is sufficient to stabilize the solution to a dS-like configuration and thus to avoid the big crunch with is typically associated with negative potentials for standard kinetic terms~\cite{Linde:2001ae, Felder:2002jk}.

\section{A cosmological landscape}
\label{sec:cosmological}

In this section we apply the techniques developed in~\cref{sec:HJ} to a complete cosmological scenario where the energy density of baryons is treated independently from DM and radiation is also included in the whole energy budget. In the following we refer to this extension as the GBI+Vbr model. Let us start by considering the two relevant components of Einstein equations:
\begin{align}
\label{eq:bar_rad_fried1}
    3H^2 &= \rho_{\mathrm{tot}} = A^{\frac{1}{1+\alpha}} \left(1- \phid^{\frac{1+\alpha}{ \alpha }}\right)^{-\frac{1}{1+\alpha}} + V(\phi) + \rho_b + \rho_r\;, \\[5pt]
    -2\phid \Hphi &= p_{\mathrm{tot}} + \rho_{\mathrm{tot}} = A^{\frac{1}{1+\alpha}}\left(1- \phid^{\frac{1+\alpha}{ \alpha }}\right)^{- \frac{1}{1+\alpha}}  \phid^\frac{1+\alpha}{\alpha}  + \rho_b + \frac{4}{3}\rho_r\;.
    \label{eq:bar_rad_fried2}
\end{align}
As explained in~\cref{sec:HJ}, the first step to express the problem in terms of the HJ formalism is to express $\dot{\phi}$ as a function of $\phi$ only. While it is possible to solve~\cref{eq:HJ_fried2} analytically for $\dot{\phi}$, it is clear that an analytical solution of~\cref{eq:bar_rad_fried2} for $\dot{\phi}$ does not exist for a general choice of $\alpha$. As a consequence, this solution has to be found numerically. In particular, once a particular expression for $H(\phi)$ is specified, a root-finding algorithm is used to compute the value of $\phid$ corresponding to a given value of $\phi$. In order to determine the evolution of a GBI+Vbr Universe in the HJ formalism, we start by determining the value of $\phi$ today directly using~\cref{eq:param} and imposing $H(\phi_0) = H_0$. The corresponding value of $\phid$ is then obtained by solving~\cref{eq:bar_rad_fried2}. As customary, $a_0$, the present value of $a$, is set to be equal to $1$. A new value (smaller for backward evolution, larger to study the future behaviour of the system) of $\phi$ is then fixed and the corresponding value for $\phid$ is determined using~\cref{eq:bar_rad_fried2}. In order to evolve the baryon and radiation's energy densities, we then integrate~\cref{eq:Hubble} to compute the value of $a$ corresponing to new evolutionary step. Finally, the corresponding value of the scalar potential $V(\phi)$ is evaluated using~\cref{eq:bar_rad_fried1}. If the steps in $\phi$ are chosen to be sufficiently small to keep numerical error under control, by iteratively repeating this procedure it is possible to compute the numerical evolution of the system. \\

\begin{figure}
    \centering
    \includegraphics[width=\textwidth]{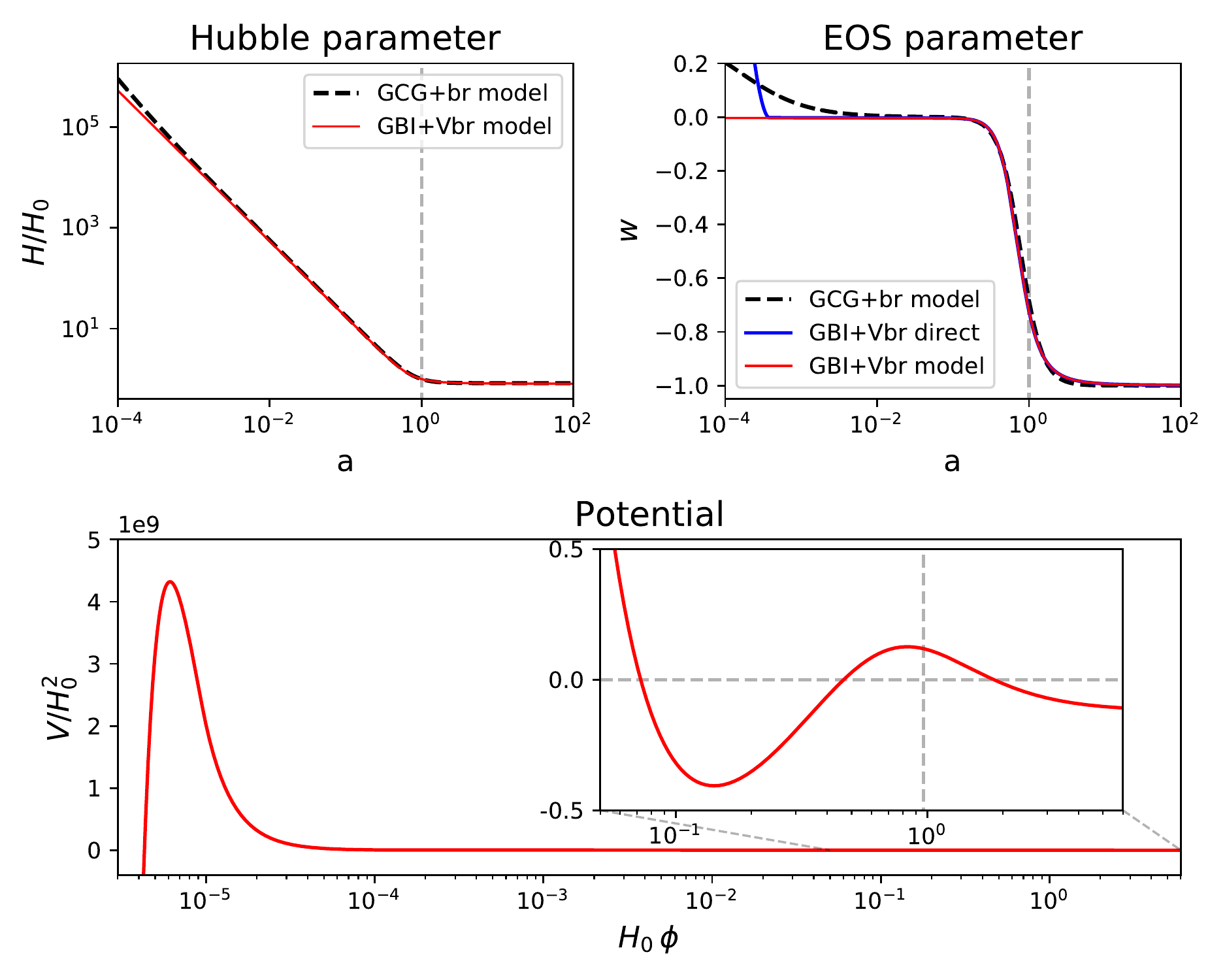}
    \caption{Top plots: comparison between a pure GBI model (black dashed line) and a model respecting~\cref{eq:param} with $\{p_i\} = \{0.669, 0.830, 2.438\}$ (red line). Independent baryon and radiation components are included in both cases. The vertical dashed lines correspond to $a=1$. The top left and right plots respectively show the evolution of $H$ and of the equation of state parameter as a function of $a$. The blue line in the top right plot is $w$ evaluated as $p_{\rm tot}/\rho_{\rm tot}$. The bottom plot shows the shape of the scalar potential as a function of $\phi$. The field range used in the bottom plot corresponds to the range used for $a$ in the two top plots.}
    \label{fig:cosm}
\end{figure}

The evolution of a GBI+Vbr Universe with Hubble parameter specified by the parameterisation of~\cref{eq:param} is plotted in~\cref{fig:cosm}, in comparison with a GCG Universe which also includes baryons and radiation. This model, which at least at late times closely resembles $\Lambda$CDM, in the following will be referred to as GCG+br. The two top plots of~\cref{fig:cosm} clearly show that the two models are in good agreement during DE and most of DM domination. This behaviour is expected, since the parameterisation in~\cref{eq:param} has been specifically chosen to capture these behaviours. However, once radiation becomes non-negligible (\emph{i.e.} when $a$ becomes smaller than $\sim 10^{-2}$), we can see that the Hubble parameters of the two models start to diverge from one another. Similarly, and more evidently, we can appreciate the divergence between the EOS parameters of the two models. The reason for this divergence is intrinsically due to the choice of specifying $H(\phi)$ as in~\cref{eq:param}. In fact, by construction, this parameterization is appropriate to describe DE and DM-like evolution, but it does not include a radiation-like behaviour in the early times. As a consequence, while the GCG+br's parameter switches to the~$H \sim a^{-2}$ behaviour typical of radiation, the GBI+Vbr model retains a~$H\sim a^{-3/2}$ behaviour which is typical of matter domination. In order to better identify the breaking of our parameterization, in the top right panel of~\cref{fig:cosm} we plot both $w$ evaluated using the left hand sides (in red) and the right hand sides (in blue) of~\cref{eq:bar_rad_fried1} and of~\cref{eq:bar_rad_fried2}. This plot clearly shows that as radiation grows the two evaluations of the same quantity differ, proving that~\cref{eq:param} has become inconsistent. \\

Finally, we conclude this section by commenting on the shape of the scalar field potential as a function of $\phi$, which is shown in the lower plot of~\cref{fig:cosm}. In the DE and in most of the DM-like phases we basically recover the same shape shown in the lower plot of~\cref{fig:scalarGCGpot}, implying that, as expected, the introduction of independent baryon and radiation components do not significantly alter the behavior of the system in its late time evolution. On the other hand, the shortcoming of the parameterization in~\cref{eq:param} is once again evident from the early time shape of $V(\phi)$ which is determined using~\cref{eq:bar_rad_fried1}. At early times, corresponding to $\phi \rightarrow 0$, $H$ grows as $1/\phi$ by construction. However, in this regime $\rho_r$ scales like\footnote{Recall that at early times $\dot{\phi} \sim 1 $ which, using~\cref{eq:Hubble}, implies $a \sim \phi^{p_1}$.} $\phi^{-4 p_1}$ which grows faster than $H^2$. As a consequence, in order for~\cref{eq:bar_rad_fried1} to be satisfied, $V(\phi)$ becomes negative to compensate the growth of the radiation component. Clearly this behavior is unphysical and it shows that the validity of~\cref{eq:param} breaks when the radiation component takes over the other energy species. 

\section{Discussion and conclusions}
\label{sec:conclusions}
In this work we have discussed the application of the HJ formalism to a slight generalization of GCG models. As we have discussed in~\cref{sec:model}, the scalar field implementation of these models is realized by a GBI theory. In order to supplement the model (which otherwise is uniquely specified by the choice of the two parameters $\alpha$ and $A$) with an additional functional freedom, we have introduced a scalar potential $V$ for the field $\phi$. We have therefore dubbed this new class of models GBI+V (further extended to GBI+Vbr in~\cref{sec:cosmological} where independent components for both baryons and radiation are included too). Interestingly, the non-standard kinetic term of the GBI action, may effectively induce a lifting which can give a dS-like solution even for a negative potential. This has some similarities with the KKLT mechanism~\cite{Kachru:2003aw, Kachru:2003sx} where a certain number of static $\overline{D3}$ branes are included in the construction to induce a lifting in the potential which is proportional to the value of the warp factor at the brane location. To recover a similar construction (which is described by the same action used in DBI inflation~\cite{Silverstein:2003hf, Alishahiha:2004eh}) an additional term proportional to the inverse of the $D3/\overline{D3}$ brane's warp factor (with a $\mp$ sign respectively) should be added to~\cref{eq:GCG_action}. \\

In~\cref{sec:HJ} we have then discussed the framing of GCG models in terms of the HJ formalism. We have first considered the $V=0$ case (\emph{i.e.} the pure GBI) and we have computed analytical approximations for the DM and DE-like asymptotic behaviours. Using these approximate solutions, we have then built a parameterization for $H(\phi)$ which is specified in terms of three parameters, denoted by $\{p_i\}$, which are respectively controlling the EOS parameter in the DM-like phase, the time at which the DM/DE transition take place and its sharpness. Moreover, we have computed the shape of the scalar potential corresponding to a given choice for $\{p_i\}$ and once again we have computed analytical approximations for the DM and DE asymptotic behaviours. In~\cref{sec:cosmological} we have shown that the HJ approach can also be used for more realistic cosmological setups, which may also include baryons and radiation. However, in general in this case it is not possible to get a closed form analytical solution and the the evolution of the system can only be computed numerically. We have shown that by construction, the parameterization of $H(\phi)$ given in~\cref{eq:param} does not allow for an early stage of radiation domination. Thus, even if radiation can be included in the model, the evolution cannot be pushed in the regime where it takes over matter. The generalization of~\cref{eq:param}, with the introduction of an additional term which could account for the early stage of radiation domination, is an interesting matter that is left for future works on this topic. Such a generalization would be a necessary step to produce a fully consistent cosmological model which could then be constrained using real data. \\

Finally, while in this work we have only focused our attention on the GCG models (and the extension we have defined), several generalizations of this model, typically specified by a modification of~\cref{eq:GCG_EOS}, exist in the literature. Some of the most famous are the so-called variable Chaplygin gas model~\cite{Guo:2005qy} (where the parameter $A$ appearing of~\cref{eq:GCG_EOS} is promoted to a $\phi$-dependent function), the new generalized Chaplygin gas model~\cite{Zhang:2004gc} (where $A$ is $a$-dependent), the modified Chaplygin gas model~\cite{Paul:2014kza} (where a term linear in $\rho$ is added in~\cref{eq:GCG_EOS}) or the Extended Chaplygin gas model~\cite{Pourhassan:2014ika} (where a series of positive powers of $\rho$ is included in~\cref{eq:GCG_EOS}). Further extensions of the present analysis which could include any of these generalizations could constitute an interesting subject for future works on this topic.

\section*{Acknowledgments}
We would like to thank Carlo Contaldi, Angelo Ricciardone and Marco Scalisi for discussions at different stages of this project. We also thank Joel Mabillard and Lukas Witkowski for very useful comments on the draft of this work. The work of M.P. was supported by STFC grants ST/P000762/1 and ST/T000791/1. M.P. acknowledges support by the European Union’s Horizon 2020 Research Council grant 724659 MassiveCosmo ERC- 2016-COG. 

\appendix
\section{Parameter Variation}
\label{sec:appendix_A}

\begin{figure}[htb]
    \centering
    \includegraphics[width=1.2\textwidth]{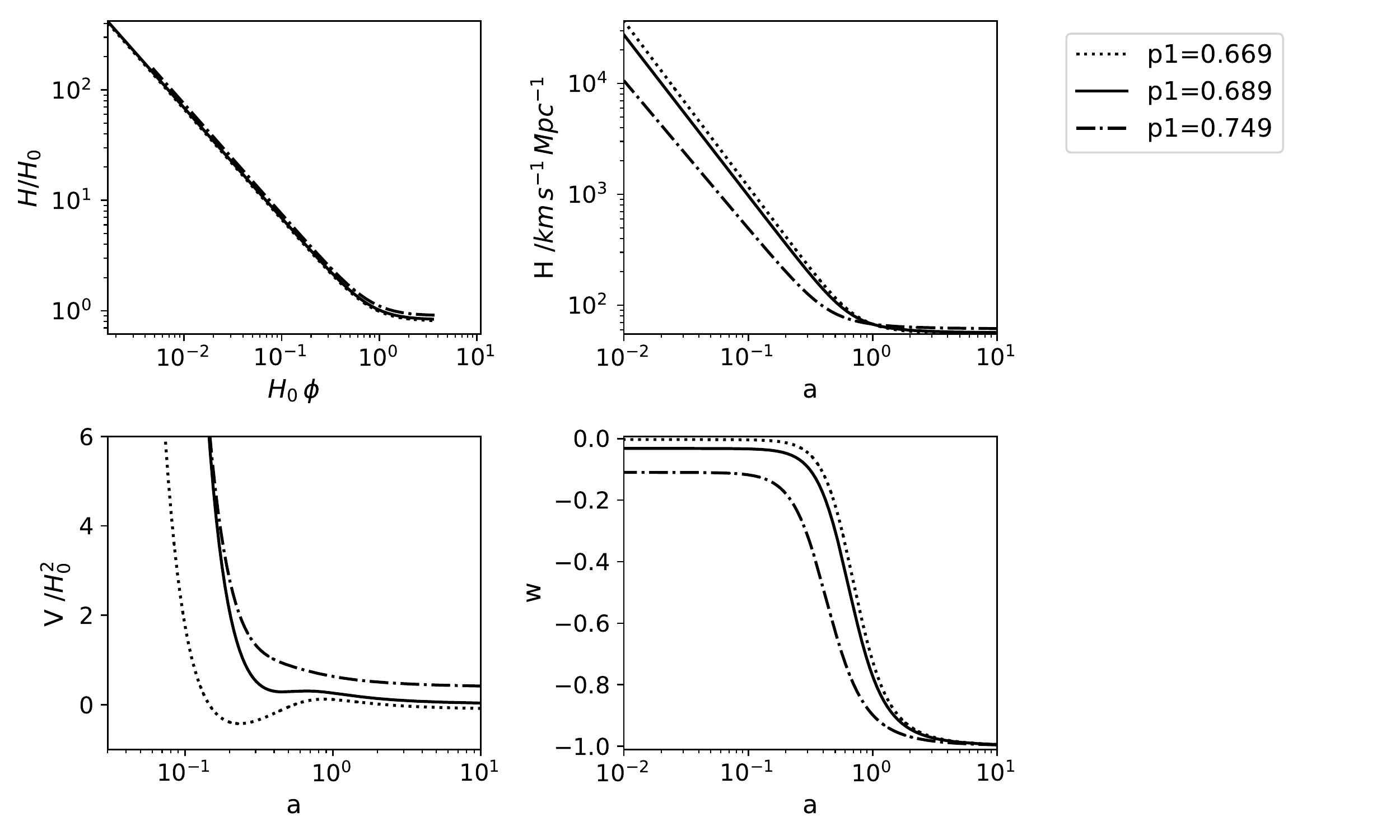}
    \caption{Impact of the variation of $p_1$ on the GCG+V model with $p_2, p_3 = 0.830, 2.438$. The top left and right plots respectively show the evolution of $H$ as a function of $\phi$ and $a$. The bottom left and bottom right plots respectively show the potential and EOS parameters as functions of $a$. It is manifest that $p_1$ controls the value of the EOS parameter of the fluid in the fast-roll regime, letting it mimick some combination of DM and DE. Larger values of $p_1$ result in steeper potentials and more negative $w$. Additionally, larger $p_1$ increases the final value of $H$ in the DE regime.}
    \label{fig:varpar_p1}
\end{figure}

\begin{figure}[htb]
    \centering
    \includegraphics[width=1.2\textwidth]{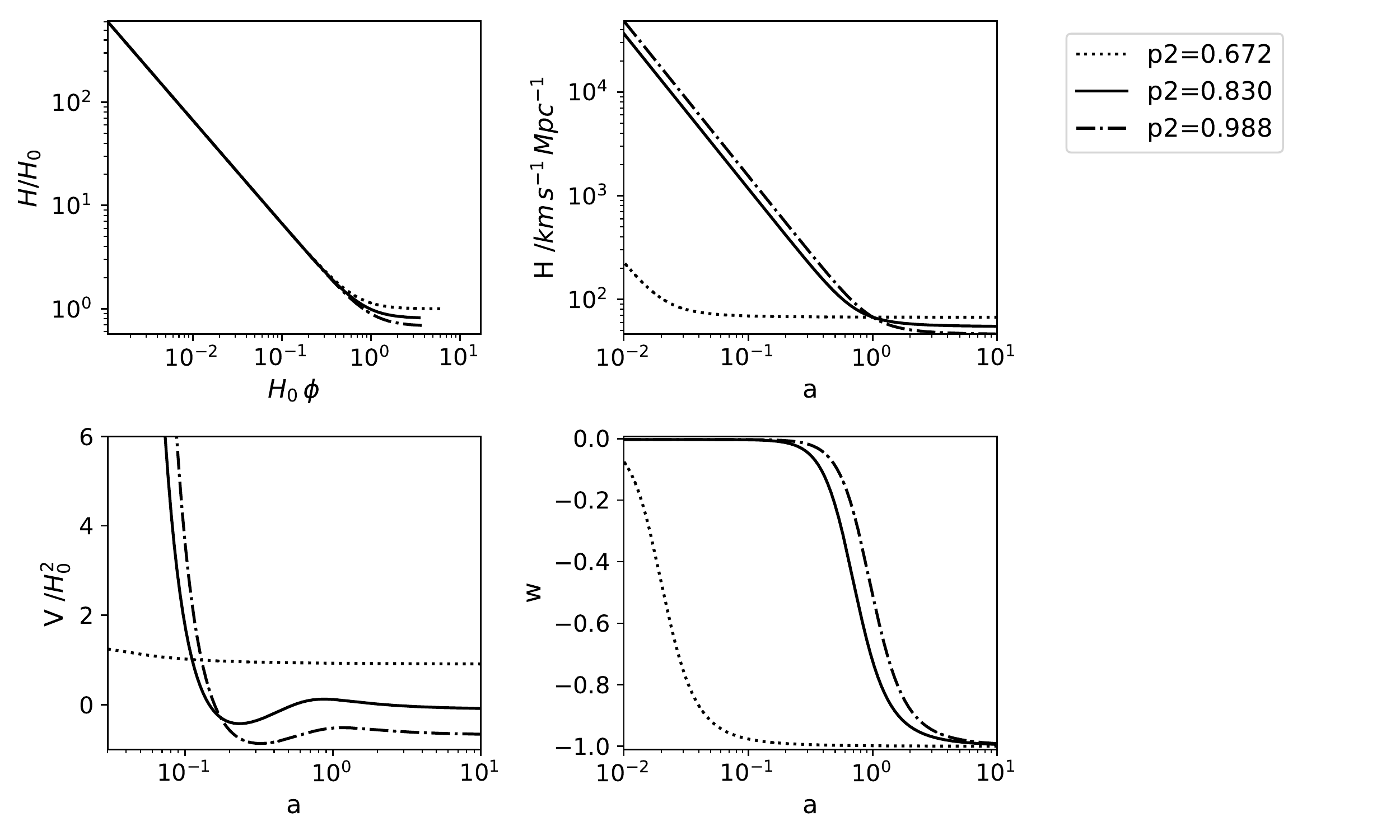}
    \caption{Impact of the variation of $p_2$ of the GCG+V model with $p_1, p_3 = 0.669, 2.438$. The top left and right plots respectively show the evolution of $H$ as a function of $\phi$ and $a$. The bottom left and bottom right plots respectively show the potential and EOS parameters as functions of $a$. $p_2$ determines the position of the DM-DE transition. The smallest value (dotted line) begins transitioning at $a\sim10^{-2}$, characterised by a flattening $H(a)$ and the $w$ moving towards $-1$. The potential has also reached its final, constant value. Additionally, larger $p_2$ decreases the final value of $H$ in the DE regime.}
    \label{fig:varpar_p2}
\end{figure}

\begin{figure}[htb]
    \centering
    \includegraphics[width=1.2\textwidth]{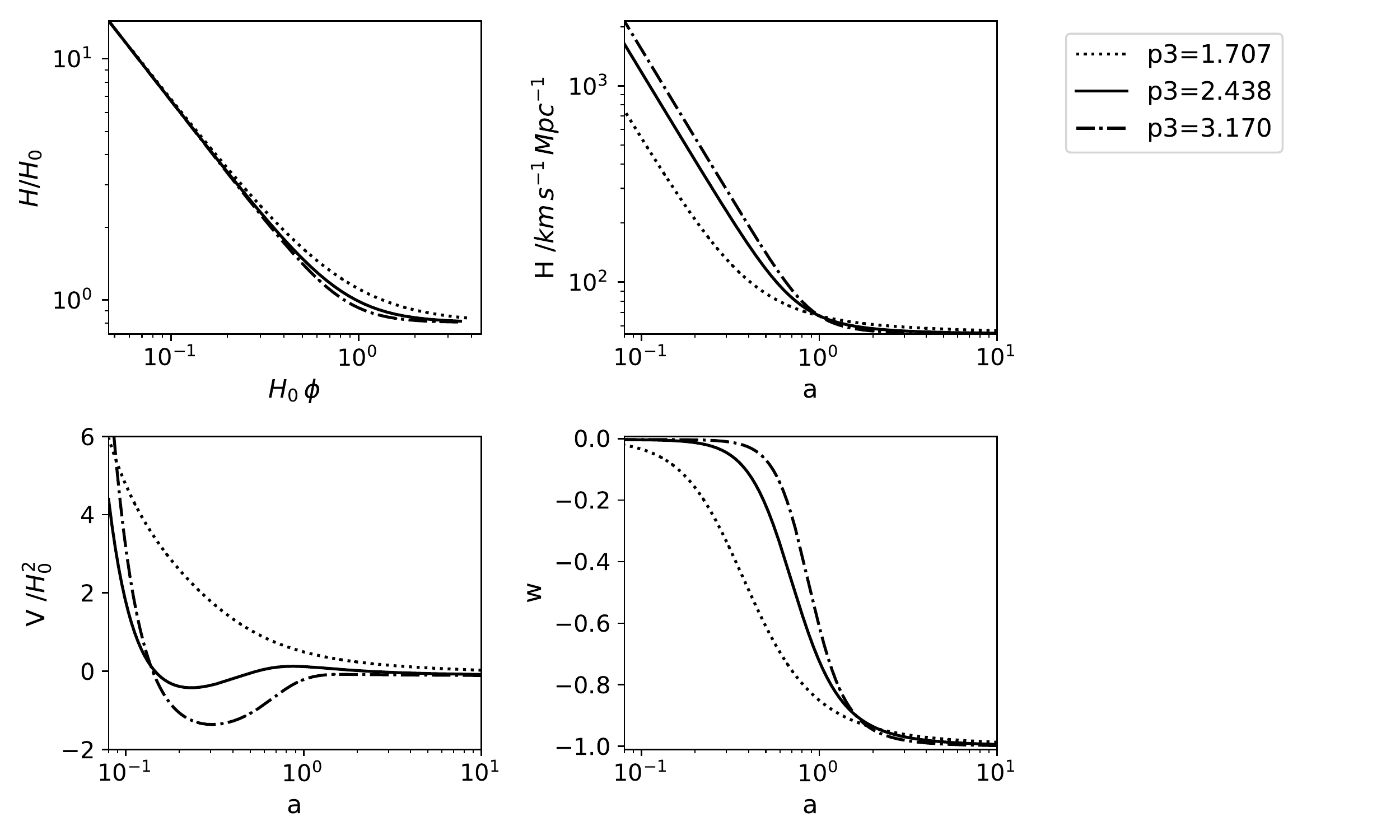}
    \caption{Impact of the variation of $p_3$ of the GCG+V model with $p_1, p_2 = 0.669, 0.830$. The top left and right plots respectively show the evolution of $H$ as a function of $\phi$ and $a$. $p_3$ determines the speed of the DM-DE transition. Lower values transition more slowly and begin doing so earlier, as seen by the difference in the sharpness of transitions of the EOS parameter (bottom-right). Higher values of $p_3$ also produce larger dips in the potential. The larger dip allows the field to keep moving quickly for longer, before reaching the uphill section where it rapidly slows down, entering the slow-roll regime.}
    \label{fig:varpar_p3}
\end{figure}

In this appendix we discuss the allowed range for the three parameters $p_1$, $p_2$ and $p_3$ as well as their impact on the cosmological evolution (respectively shown in~\cref{fig:varpar_p1},~\cref{fig:varpar_p2} and~\cref{fig:varpar_p3}). For this purpose, each of the three parameters is varied individually, while keeping the other two fixed at the values presented in~\cref{sec:HJ}.\\

We start our treatment by discussing the model's viable parameter space which is constrained by two requirements; firstly, for a given parameter choice there must exist a value $\phi_0$ such that $H(\phi_0) = H_0$. Since $H(\phi)$ is monotonically decreasing and for $\phi\rightarrow\infty$ it goes to $H\rightarrow H_0 \,p_1/p_2$, we can immediately conclude that $p_2 \geq p_1$ is required for a solution to $H(\phi_0) = H_0$ to exist. Secondly, to prevent an unbounded negative potential in the past, the first term must be larger in magnitude than the second one. From~\cref{eq:V_small_phi} it is trivial to show that this corresponds to
\begin{equation}
    3p_1 \geq 2
    \left[1 + A\left(\frac{2p_1}{\phi^2}\right)^{-(1+\alpha)}\right]^{1+\alpha} \;,
\end{equation}
which reduces further to $p_1 \in [2/3, \infty)$ as $\phi \rightarrow 0$, setting the second constraint. \\

We proceed by discussing the variation in the behavior of the model induced by the three parameters $p_1$, $p_2$ and $p_3$. Let us start by focusing on $p_1$. In the fast-roll regime, the model imitates a mixture of DM and DE, whose relative proportions are governed by the parameter $p_1$. Smaller values correspond to steeper potentials, producing a faster-rolling field with EOS parameters closer to zero (\cref{fig:varpar_p1}). On the contrary, larger values produce shallower potentials and slower rolling fields, resulting in negative EOS parameters. More formally, this can be shown by using~\cref{eq:Hubble} which, given the fast-roll at early times, reduces to $H \simeq  \textrm{d} \ln a / \textrm{d} \phi$. Since at early times we also have $H \simeq p_1 / \phi$, we can easily get $a\propto \phi^{p_1}$, which in turn implies $H(a)\sim a^{-1/p_1}$. This behaviour is evident in the $H(a)$ plot of~\cref{fig:varpar_p1}. Therefore, by increasing the values of $p_1$, we approach the behaviour of DE, which is given by $H\sim a^0$ and is reached in the limit $p_1 \rightarrow \infty$. \\

The point in time at which the DM-DE transition occurs is governed by the parameter $p_2$. This parameter translates the curve $H(a)$ (\cref{fig:varpar_p2}), keeping both the shape and $H_0$ fixed. Smaller values result in earlier transitions; in particular the dotted line of~\cref{fig:varpar_p2} is an extreme example. The early transition beginning at $a \sim 10^{-2}$ is evident from the bottom right panel, along with the potential which has already flattened (bottom left panel), producing slow-roll behaviour. The ratio between $p_1$ and $p_2$ also controls the final value of the Hubble parameter in DE-domination. This is evident from the top two plots in~\cref{fig:varpar_p1} and in~\cref{fig:varpar_p2}. \\

Finally, the rate of the DM-DE transition is controlled by the parameter $p_3$, whose impact is clearly evident from the bottom right plot of~\cref{fig:varpar_p3}, which shows a clear change in the rate at which the EOS parameter switches from $0$ to $-1$. Additionally, larger values of $p_3$ result in deeper dips in the potential. These cause the field to keep rolling quickly until the last minute, when the potential suddenly sharply rises, forcing the field to rapidly slow down and enter the slow-roll regime.

\bigskip

\bibliographystyle{JHEP}
\bibliography{biblio}

\end{document}